\def\draftversion{false}
\RequirePackage{ifthen}
\ifthenelse{\equal{\draftversion}{true}}{
  \documentclass[aps,prb,10pt,galley,amsmath,amssymb,showpacs,letterpaper]{revtex4}
}{
  \documentclass[aps,prb, 10pt,twocolumn,amsmath,amssymb,showpacs,letterpaper]{revtex4}
}

\usepackage{graphicx}
\usepackage{epsfig}
\usepackage{dcolumn}
\usepackage{bm}
\usepackage{subfigure}
\usepackage{hyperref}

\hypersetup{backref=true, pdfnewwindow=true, colorlinks=true,linkcolor=blue, anchorcolor=blue, citecolor=blue, filecolor=blue, menucolor=blue, urlcolor=blue}

\usepackage[usenames,dvipsnames]{color}

\usepackage{ulem} 

\ifthenelse{\equal{\draftversion}{true}}{
  \marginparwidth 2.7in
  \marginparsep 0.5in
  \newcounter{comm} 
  \def\commnext{\stepcounter{comm}}
  \def\commtext{{\bf\color{blue}[\arabic{comm}]}}
  \def\commmar{{\bf\color{blue}[\arabic{comm}]}}
  \def\dvm#1{\commnext\marginpar{\small DV\commmar: #1}\commtext}
  \def\mtm#1{\commnext\marginpar{\small MT\commmar: #1}\commtext}
  \def\mlab#1{\marginpar{\small\bf #1}}
  
}{
  \def\dvm#1{}
  \def\mtm#1{}
  \def\kgm#1{}
  \def\mlab#1{}
  
}

\def\beq{\begin{equation}}
\def\eeq{\end{equation}}
\newcommand{\equ}[1]{Eq.~(\ref{eq:#1})}

\def\k{\textbf{k}}

\def\Tr{\textrm{Tr}}

\def\zval{\bar{z}}

\begin{document}

\title{Adiabatic Pumping of Chern-Simons Axion Coupling}

\author{Maryam Taherinejad}
\email{mtaheri@physics.rutgers.edu}

\affiliation{Department of Physics and Astronomy, Rutgers University,
Piscataway, New Jersey 08854-0849, USA}

\author{David Vanderbilt}

\affiliation{Department of Physics and Astronomy, Rutgers University,
Piscataway, New Jersey 08854-0849, USA}

\date{\today}

\pacs{03.65.Vf,73.20.At,73.43.-f,71.15.Rf,75.85.+t}

\begin{abstract}
We study the adiabatic pumping of the Chern-Simons axion (CSA)
coupling along a parametric loop characterized by a non-zero
second Chern number $C^{(2)}$ from the viewpoint of the hybrid
Wannier representation, in which the Wannier charge centers
(WCCs) are visualized as sheets defined over a projected 2D
Brillouin zone.  We derive a new formula for the CSA coupling,
expressing it as an integral involving
Berry curvatures and potentials defined on the WCC sheets.
We show that a loop characterized by a non-zero $C^{(2)}$
requires a series of sheet-touching events at which $2\pi$
quanta of Berry curvature are passed from sheet to sheet, in
such a way that $e^2/h$ units of CSA coupling are pumped by a
lattice vector by the end of the cycle.  We illustrate these
behaviors via explicit calculations on a model tight-binding
Hamiltonian and discuss their implications.

%

\end{abstract}

\maketitle


The response of materials to electromagnetic fields has been the
subject of studies for many years, both for reasons of fundamental interest
and for technological applications. The discovery of topological
insulators (TIs) and related classes
of materials in recent years has generated interest
in the Chern-Simons axion (CSA) coupling, which makes an isotropic
contribution $\alpha^{CS}$ to the magnetoelectric response tensor
of the material.  This coupling is conventionally expressed in
terms of a dimensionless parameter $\theta$ defined via
\begin{equation}
\alpha_{ij}^\textrm{CS}=\dfrac{\theta e^2}{2\pi h} \delta_{ij},
\label{eq:CS_theta}
\end{equation}
where $\theta$ is determined by the band structure of the insulator
via an integral over the Brillouin zone (BZ) of a Chern-Simons
3-form according to
\begin{equation}
\theta=-\dfrac{1}{4\pi} \int d^3k \,
\epsilon^{ijk}\,\Tr[A_i \partial_j A_k- i\dfrac{2}{3}A_i A_j A_k].
\label{eq:theta}
\end{equation}
Here $A^{nm}_i=i\langle u_{n} \vert \partial_i \vert u_{m}
\rangle$ is the Berry connection (or non-Abelian gauge field)
in Cartesian direction $i$, where $u_n(\mathbf{k})$ is the periodic
part of the Bloch function of the $n$'th occupied band, and the
trace is over occupied bands.

The ground-state properties of a band insulator are
invariant under any gauge transformation, that is, any unitary
transformation $U_{nn'}(\mathbf{k})$
that mixes only the occupied bands.
It can be shown that an arbitrary gauge transformation either
leaves the 3-form integral in \equ{theta} unchanged or else shifts
it by exactly $2\pi$ times an integer.  Thus, $\theta$ is best regarded
as a phase angle that is only well-defined modulo $2\pi$.
As a consequence, the presence of either time reversal (TR) or inversion
(either of which flips the sign of $\theta$) requires $\theta$ to be
quantized to an integer multiple of $\pi$, with an odd/even value
corresponding to an odd/even strong $Z_2$ topological
index of a TR-invariant 3D insulator.\cite{essin-prl09, qi-prb08}
One way to understand the ambiguity of $\theta$ modulo $2\pi$,
which corresponds to an ambiguity of $\alpha^\textrm{CS}$
modulo $e^2/h$, is to realize that the magnetoelectric coupling
is related to the surface anomalous Hall conductivity (AHC) by
$\sigma =(\theta/2\pi+C)e^2/h$.
Thus while $\theta$ can be calculated from the bulk band
structure, the measurable magnetoelectric response can
be changed by a quantum if a layer with non-zero Chern
number is attached to the surface of material, changing
the effective value of $\theta$ by $2\pi$.

An interesting consequence of this $2\pi$ ambiguity is
that if an insulator is allowed to evolve adiabatically around
a closed loop in the space of parameters determining the
crystal Hamiltonian, with the gap remaining open,
then the fact that the system returns to the initial physical
state means that $\theta$ must either return to its
original value or change by an integer multiple of $2\pi$,
where the integer $C^{(2)}$ is known as a ``second Chern number.''
This possibility of ``pumping $\theta$ by $2\pi$''
has been discussed and demonstrated for some theoretical
models,\cite{essin-prl09,qi-prb08} but the characteristic
behaviors of a system undergoing such an adiabatic loop have
largely remained unexplored.

Recently, we have shown that the hybrid Wannier representation
can be a useful and insightful tool for computing topological
indices and inspecting the topological properties of 3D
insulators.\cite{taheri-prb14}
In this approach, the occupied-state wavefunctions are transformed
into a maximally-localized Wannier representation in one chosen
Cartesian direction, while remaining Bloch-like in the orthogonal
directions.  The resulting hybrid Wannier functions (HWFs)
inherit the topological character of the insulator, and plots of
their Wannier charge centers (WCCs) over the 2D BZ (``Wannier
sheets'') were shown to provide a useful means of visualizing
the topological properties of insulators, allowing to discriminate
immediately between normal, strong topological, weak topological,
crystalline topological, and related 
states.\cite{taheri-prb14,soluyanov-prb11,Alexandradinata-arx12,Alexandradinata-arx14}

With these motivations, we ask what happens if an adiabatic
cycle that pumps $\theta$ by $2\pi$ is viewed from the point of
view of the HWF representation.  How do the WCC sheets evolve?
Is there a characteristic behavior of this evolution of the WCC
sheets that signals the presence of a non-trivial cycle (i.e.,
a non-zero second Chern number)?  We find that there is indeed
such a characteristic behavior.  Specifically, we show that quanta
$e^2/h$ of Berry curvature are passed from one WCC sheet to the next
in a series of isolated band-touching events, in such a way that
one quantum of Berry curvature is pumped by an entire lattice
vector by the close of the cycle.  We illustrate this amusing and
instructive result via numerical calculations on a 3D spinor
tight-binding model and discuss its implications.

We begin with a brief review of the construction of the hybrid
Wannier representation.  We choose a special direction, here $\hat{z}$,
along which the Wannier transformation is carried out, so that
the HWFs are localized in $z$ while remaining Bloch-like in the
other two directions.\cite{sgiarovello-prb01,marzari-prb97}
Explicitly,
\begin{equation}
\vert W_{ln}(k_x,k_y)\rangle =
\dfrac{c}{2\pi}\int dk_z
e^{i\mathbf{k}\cdot ({\mathbf{r}} -lc
\hat{z})} \vert u_{n,\mathbf{k}} \rangle,
\label{eq:hwf}
\end{equation}
where $l$ is a layer index and $c$ is the lattice constant
along $\hat{z}$. In general, there is a $U(N)$ gauge freedom
in choosing the $N$ representatives of the occupied space,
$\vert \tilde{u}_{n,\mathbf{k}} \rangle=\sum_m U_{nm}
\vert u_{m,\mathbf{k}} \rangle$, but there is a unique
gauge that minimizes the spread functional of the
WFs along $\hat{z}$.\cite{marzari-prb97}
These maximally localized HWFs and their WCCs
$\zval_n(k_x,k_y)=
\langle W_{n0}\vert z \vert W_{n0} \rangle$
can be constructed using standard methods.\cite{marzari-prb97,wu-prl06}

For a 2D insulator the WCCs can be plotted as curves $\zval_n$
vs.\ $k_\perp$ in a 1D projected BZ,\cite{soluyanov-prb11,yu-prb11}
while for a 3D insulator they can be visualized as sheets plotted
over the 2D projected BZ. In previous work\cite{taheri-prb14}
we have shown that these WCC sheets provide an insightful
characterization of the topological character of the insulator
in question, allowing one to see how electrons are adiabatically
pumped along $\hat{z}$ as $k_x$ and $k_y$ are varied.
For example, in a 2D Chern insulator
the WCCs shift by one or more lattice constants
along $z$ as $k_x$ evolves across the projected BZ,
pumping units of charge along that direction. This extra charge
is removed from the edge as the edge band crosses the Fermi energy.
Time reversal invariant (TRI) insulators have zero Chern numbers 
but are characterized by $Z_2$ topological indices that are also 
reflected in the structure of WCC sheets.
For example, a 3D TRI insulator is characterized by one strong
and three weak $Z_2$ indices, which can be
determined by examining how the WCC sheets connect along TRI
lines in the projected BZ for different Wannierization
directions.\cite{taheri-prb14}

It is also of interest to consider the behavior of the
WCC sheets as the crystal Hamiltonian is carried adiabatically
around a loop defined by some cyclic parameter $\alpha$
corresponding, e.g., to some combination of
atomic displacements and/or external fields. A celebrated
result of Thouless\cite{Thouless-prb83}
is that this results in quantized adiabatic charge transport, i.e.,
the pumping of exactly one electron per unit cell by a lattice
vector ${\bf R}$ during the cycle.  Normally ${\bf R}\!=\!\mathbf{0}$,
but for example if ${\bf R}=c\hat{z}$ this corresponds to the pumping
of one electron by one period along $z$ during the cycle
(a first Chern number of $C\!=\!1$), i.e., a change
in electric polarization $\Delta P_z=-e/A_{\rm cell}$ with $A_{\rm
cell}$ the projected unit cell area.

Let us see how this evolution occurs from the viewpoint of the HWF
representation.  Intuitively, we expect each WCC sheet to drift
along $z$ with increasing $\alpha$ such that it replaces the one
above it, and is replaced by the one below it, at the end of the
cycle.  We begin by defining Berry potentials ``living on the
sheets'' representation as
\begin{equation}
A_{x,ln,l'm}=\langle W_{ln} \vert i\partial_x \vert W_{l'm} \rangle \,,
\label{eq:Axdef}
\end{equation}
and similarly for $A_y$.  These are functions of $(k_x,k_y)$ and also
matrices in the space of sheet labels $ln$ (the $n^{\rm th}$ sheet in
cell $l$ along $z$).  The corresponding Berry potentials in the Bloch
representation are then just
\begin{eqnarray}
&& A_{x,nm}(\k)=\sum_l e^{ik_zlc}A_{x,0n,lm}(k_x,k_y) \,,
\label{eq:AjAj}\\
&& A_{z,nm}(\k)=\zval_n(k_x,k_y)\,\delta_{nm} \,.
\label{eq:Azz}
\end{eqnarray}
Plugging into the Berry-phase formula for the electronic contribution
$P_j=-e(2\pi)^{-3} \sum_n\int d^3k\,A_{jn}(\k)$,
we find
\begin{eqnarray}
&&P_j=\frac{-e}{(2\pi)^2c}\sum_n\int d^2k\;A_{j,0n,0n}(k_x,k_y) \,,
\label{eq:Pj}\\
&&P_z=\frac{-e}{(2\pi)^2c}\sum_n\int d^2k\;\zval_n(k_x,k_y) \,,
\label{eq:Pz}
\end{eqnarray}
where $j=\{x,y\}$ in Eq.~(\ref{eq:Pj}).  For the case of a parametric
loop that pumps electrons along $z$, the change $\Delta P_z=-e/A_{\rm cell}$
would occur via the gradual migration of the $\zval(k_x,k_y)$ along the
$+\hat{z}$ direction, with a relabeling of sheets required at the
end of the loop.


Now we again consider an adiabatic cycle in a
3D insulator, but instead of resulting in the pumping of charge,
we examine the case where it results in the pumping of the
CSA coupling, increasing $\theta$ by $2\pi N$
where $N$ is a nonzero integer (second Chern number $C^{(2)}=N$).
This corresponds to a pumping of Berry curvature, instead of
electric charge, along $z$ during the adiabatic cycle.
For this purpose
we now extend the above discussion by defining a Berry curvature
on the WCC sheets as
$\Omega_{xy,ln,l'm}(k_x,k_y)=
  i\langle\partial_x W_{ln}\vert \partial_y W_{l'm}\rangle
 -i\langle\partial_y W_{ln}\vert \partial_x W_{l'm}\rangle$.
The relation to the Berry curvature in the Bloch representation
is similar to that for $A$ in \equ{AjAj}, namely
\begin{equation}
\Omega_{xy,nm}(\k)=\sum_l e^{ik_zlc}\Omega_{xy,0n,lm}(k_x,k_y) \,.
\label{eq:OO}
\end{equation}
The intrinsic AHC $\sigma_{yx}$ of the
crystal is just given by integrating the trace of $\Omega_{xy}$
in the Bloch representation over the 3D BZ, and this is easily shown
to be equal to
$(e^2/hc)\sum_n C_n$ where $C_n$ is the Chern number of the $n^{\rm th}$
sheet in the home unit cell, given by
$C_n=(2\pi)^{-1}\int d^2k \, \Omega_{xy,0n,0n}$. We shall exclude
quantum anomalous Hall insulators from our discussion here, so
we can assume that $\sum_nC_n=0$, but importantly the individual
$C_n$ can be nonzero.

We now address the central issue of this Letter, namely, how to
represent the CSA coupling $\theta$ in the HWF representation.
Starting from \equ{theta}, this can be written as
\begin{equation}
\theta=\theta_{z\Omega}+\theta_{\Delta xy}
\label{eq:theta-sum}
\end{equation}
where
\begin{equation}
\theta_{z\Omega}= - \frac{1}{2\pi}\int d^3k\, \Tr[A_z\Omega_{xy}] \,,
\label{eq:theta_zo_3d}
\end{equation}
\vspace{-0.4cm}
\begin{equation}
\theta_{\Delta xy}=- \frac{1}{2\pi}\int d^3k\,
  \Tr[A_y \partial_z A_x-iA_z[A_x,A_y]]
\label{eq:theta_dxy_3d} \,.
\end{equation}
Performing the $k_z$ integrals with the aid of Eqs.~(\ref{eq:AjAj}),
(\ref{eq:Azz}) and (\ref{eq:OO}), these are expressed in the HWF
representation as
\begin{equation}
\theta_{z\Omega}=-\frac{1}{c}\int d^2k\sum_n \zval_n \, \Omega_{xy,0n,0n} \,,
\label{eq:theta_zo}
\end{equation}
\vspace{-0.4cm}
\begin{equation}
\theta_{\Delta xy}=\frac{i}{c}\int d^2k\sum_{lmn} (\zval_{lm}-\zval_{0n})
  A_{x,0n,lm}A_{y,lm,0n} \,,
\label{eq:theta_dxy}
\end{equation}
where $\zval_{lm}=lc+\zval_m$.
In deriving \equ{theta_dxy} we have used that
\begin{equation}
\dfrac{c}{2\pi} \int dk_z \Tr[A_y \partial_z A_x]=\sum_l \sum_{nm}
(ilc)A_{x,0n,lm}A_{y,lm,0n} \,.
\label{eq:theta_ydx}
\end{equation}
Eqs.~(\ref{eq:theta-sum}) and (\ref{eq:theta_zo}-\ref{eq:theta_dxy})
constitute a major result of the present work.
Note that the integrands in Eqs.~(\ref{eq:theta_zo}-\ref{eq:theta_dxy})
are gauge-invariant in the Bloch directions, i.e., unchanged under
a phase twist $\exp(i\varphi_n(k_x,k_y))$.

Of primary concern to us here is the
``Berry curvature dipole'' term $\theta_{z\Omega}$ in
\equ{theta_zo}, which describes
the extent to which concentrations of positive and negative Berry
curvature on the WCC sheets, given by $\Omega_{xy,0n,0n}(k_x,k_y)$,
are displaced from one another along the $\hat{z}$ direction as given by
$\zval_n(k_x,k_y)$.
Note that $\theta_{z\Omega}$ is shifted by $-2\pi C_n$
if the choice of WCC sheets comprising the home unit
cell is changed so as to shift some $\bar{z}_n$ by $c$.
The $\theta_{z\Omega}$ term is therefore the one
that has the $2\pi$ ambiguity, and
we shall see that it is responsible for the pumping of CSA coupling
in a non-trivial adiabatic cycle.
The second term $\theta_{\Delta xy}$, given by \equ{theta_dxy},
is a kind of intersheet contribution to $\theta$, in which the
$z$-separation between sheets at $(k_x,k_y)$ is coupled to the
off-diagonal (inter-sheet) matrix elements of the Berry potentials.
There is no $2\pi$ ambiguity associated with
this term, and as we shall see, it typically remains small
even when $\theta$ is at or near $\pi$.  We regard it as a
non-topological correction term that is needed for quantitative
accuracy but does not otherwise play a role in understanding
the topology of an insulator or the evolution of its CSA
coupling.

We now illustrate the concepts introduced above in the context of
a simple tight-binding model.
Following Essin
\textit{et al.},\cite{essin-prl09} we start with the Fu-Kane-Mele (FKM)
model,\cite{fu-prl07} which is a four-band model of
$s$ orbitals on a diamond lattice with spin-orbit interaction,
\begin{equation}
H_\textrm{FKM}=\sum_{<ij>}t(\textbf{e}_{ij})\,c_i^\dagger c_j + i
\lambda_{\rm so}
\sum_{\ll ij \gg} c_i^\dagger \textbf{s} \cdot
( \textbf{d}_{ij}^1 \times \textbf{d}_{ij}^2) c_j \, .
\label{eq:3d-km}
\end{equation}
The first term is a sum over first-neighbor hoppings, where
$\textbf{e}_{ij}$ is the bond vector, while the second term
involves second-neighbor hops in which vectors
$\textbf{d}_{ij}^{1,2}$ describe the two
first-neighbor bonds that make up the second-neighbor hop.
We take the cubic lattice constant to be unity.
In the original FKM model $t(\textbf{e}_{ij})=t_0$ independent of
hopping direction, but following Ref.~\onlinecite{essin-prl09} we
take $t(\textbf{e}_{ij})=t_0(3+\delta)$ for the bond along
(111) and $t_0$ for the other three bonds.
We set the first-neighbor and spin-dependent second-neighbor
hoppings to $t_0=1$ and $\lambda_{\rm so}=1$ respectively, and
assume two bands are occupied.

Starting from cubic symmetry at $\delta=-2$, where there
are gap closures at the high-symmetry $X$ points,
the symmetry is lowered and the gap is opened as $\delta$ is
varied, allowing trivial, weak, and strong topological phases to
be accessed.\cite{taheri-prb14}
The strong topological and trivial phases are separated from
each other by a band touching at the $\Gamma$ point
when $\delta=0$.
Again following Essin \textit{et al.}\cite{essin-prl09} we add a staggered
Zeeman field $h$, and define an adiabatic loop parametrized by
$\delta(\alpha)=m\,\cos(\alpha)$
and $h(\alpha)=m\,\sin(\alpha)$ where $\alpha$ runs from 0 to $2\pi$,
such that the system remains insulating on the loop and $\theta$ is
pumped by $2\pi$.
The HWF representation is constructed with $\hat{z}$
along the (111) direction, parallel to the bond with altered
hopping strength.

\begin{figure}
\includegraphics[width=3.4in]{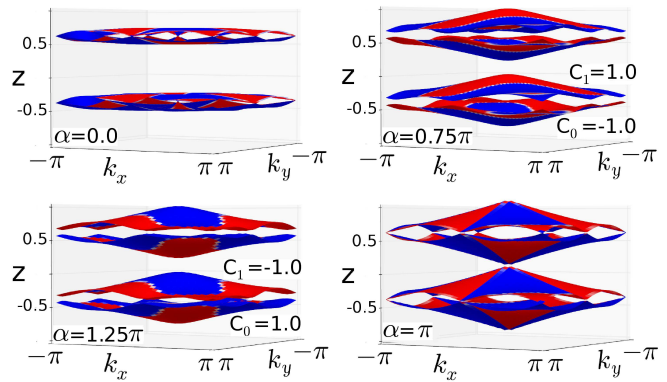}
\caption{\label{fig:bzpump}
The two WCC sheets of the half-filled FKM model, and one set
of periodic images, at four stages
$\alpha=(0,\,3\pi/4,\,\pi,\,5\pi/4)$ along the parametric cycle
(clockwise from upper left).
Blue and red colors show positive and negative values 
of Berry curvature $\Omega_z$ on the sheets. 
The Chern numbers associated with the individual
WCC sheets are shown for those cases where sheets do not touch.}
\end{figure}

\begin{figure}
\includegraphics[width=2.9in]{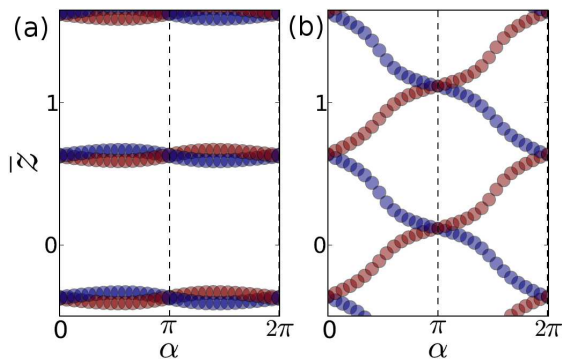}
\caption{\label{fig:curvature_pump}
The WCCs at the (a) $M$-point and (b) $\Gamma$-point
as they evolve around the adiabatic loop. 
Blue and red colors show positive and negative values 
of Berry curvature at these points.}
\end{figure}

The WCC sheets derived from the two occupied bands in the FKM model
are shown in Fig.~\ref{fig:bzpump}, where one pair of sheets
and one copy of their periodic images along $\hat{z}$ are shown
for some points around the adiabatic loop. The evolution of the
Wannier sheet positions
at the four TRI points, namely at $\Gamma$ and at the three equivalent
$M$ points, is plotted in Fig.~\ref{fig:curvature_pump}.

The system has TR symmetry at $\alpha\!=\!0$ and
$\pi$, where the system is $Z_2$-even and $Z_2$-odd respectively,
and where the WCC sheets pair up at the four TRI-points
due to Kramers degeneracy.\cite{taheri-prb14}
In the normal phase at $\alpha\!=\!0$ this results in a pair of sheets
connected by Dirac points at all four TRI momenta, and each pair is
well separated from its neighbors along $\hat{z}$.  As $\alpha$
increases, the Dirac crossings are gapped and the sheets begin
to separate.  At the three $M$ points the separation
between the pair remains quite small, and the same sheets touch
again at $\alpha\!=\!\pi$, as is obvious from
Fig.~\ref{fig:curvature_pump}(a).  At the $\Gamma$ point, however, the
behavior is much more interesting; the sheets separate strongly
and eventually reconnect with their neighbors from the next unit
cell along $\hat{z}$ when $\alpha\!=\!\pi$.  The swapping of
partners at an odd number of the TRI points (here, only at $\Gamma$)
is characteristic of the strong topological ($Z_2$-odd) phase at
$\alpha\!=\!\pi$.  Note, however, that the WCC sheets, taken together,
have no net displacement along the $\hat{z}$ direction, so no charge
is pumped.

Now we ask what happens to the CSA coupling $\theta$ during this cycle,
and to do this we inspect the Berry curvature $\Omega_{xy}$ on the
sheets. This is represented by the color-scale shading
in Figs.~\ref{fig:bzpump} and \ref{fig:curvature_pump}.
Recall that the evolution of $\theta$
is expected to be reflected in the behavior of the $\theta_{x\Omega}$
term as given by Eq.~(\ref{eq:theta_zo}).  We immediately see
that the behavior near the $M$ points is uninteresting; positive
and negative Berry curvature contributions separate slightly at
first, but they then reverse and recross, and never give a large
contribution to $\theta_{z\Omega}$.

\begin{figure}
\includegraphics[width=2.9in]{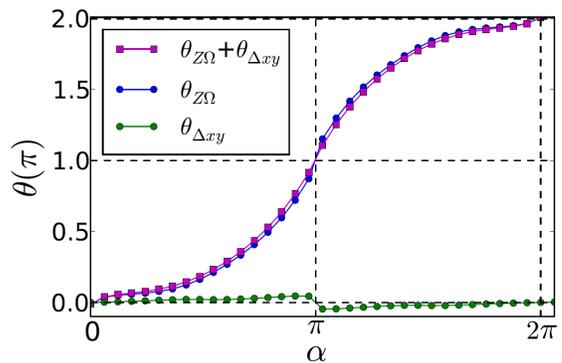}
\caption{\label{fig:theta}
The CSA coupling $\theta(\alpha)$, and the contributions of the
topological term $\theta_{z\Omega}$ and the correction term
$\theta_{\Delta xy}$, for the FKM model as it evolves around the
adiabatic loop.}
\end{figure}

Near $\Gamma$, however,
the story is strikingly different.  A
negative
(red) increment of Berry curvature is transported along $+\hat{z}$ while a
positive (blue) contribution is carried along $-\hat{z}$ as $\alpha$ evolves
from 0 to $\pi$. 
For small and positive $\alpha$ we intuitively
expect that the total Berry curvature near $\Gamma$ in the
top and bottom sheets (at $\bar{z}_2$ and $\bar{z}_1$)
should be $-\pi$ and $\pi$ respectively,
characteristic of a weakly gapped Dirac point. Thus the contribution
to the Berry curvature dipole term $\theta_{z\Omega}$ from the vicinity
of $\Gamma$, which is approximately
$\pi(\bar{z}_2(0,0)-\bar{z}_1(0,0))/c$, grows gradually as $\alpha$
increases and the sheets get further apart at $\Gamma$.  As
$\alpha\rightarrow\pi$ the separation between the sheets at $\Gamma$
approaches a full lattice constant $c$ and the contribution to
$\theta_{z\Omega}$ approaches $\pi$.  This expectation is confirmed in
Fig.~\ref{fig:theta}, where we plot the evolution of $\theta$
\textit{vs.}\ $\alpha$, and also its two individual contributions
$\theta_{z\Omega}$ and $\theta_{\Delta xy}$, as computed from
Eqs.~(\ref{eq:theta_zo}-\ref{eq:theta_dxy}).
Compared to $\theta_{z\Omega}$, the non-topological $\theta_{\Delta xy}$
term is almost negligible everywhere around
the adiabatic loop, with the possible exception of the vicinity of
the $Z_2$-odd phase where it reverses suddenly, as a result of the
WCC sheets from adjacent layers coming close to each other at
the $\Gamma$-point. 

As $\alpha$ passes through $\pi$ there is a Dirac touching at $\Gamma$
between sheet 2 in the home cell and sheet $1$ in the cell above,
with a hand-off of $-2\pi$
units of Berry curvature (or a Chern number of $-1$) from the former
to the latter, and with the concentration of Berry curvature near
$\Gamma$ in band 2 switching from $-\pi$ to $\pi$.
A direct evaluation of \equ{theta_zo} would show $\theta_{z\Omega}$
and $\theta$ dropping discontinuously by $2\pi$ as
$\alpha$ crosses through $\pi$, but we make use of the gauge freedom
to apply a $2\pi$ shift of $\theta$ to impose physical continuity
when drawing the curves in Fig.~\ref{fig:theta}.

Of course, here we have illustrated the behavior of just one model
system, and we have found that the pumping of $\theta$ by $2\pi$
is accomplished by a series of touching events between WCC sheets,
such that one Chern number of Berry curvature is handed off to the
neighboring sheet with each touching.  But having seen this, it is
now clear in retrospect that {\it any cycle that pumps $\theta$ by
$2\pi$ must involve such a sequence of touching events.} For, if these
events did not occur, the CSA coupling could not be passed along by
a lattice vector during the cycle.  Incidentally, this observation
also explains why a non-trivial $\theta$ pumping cycle is impossible
in a system with a single occupied band, since in this case the WCC
sheets are always separated by $c\hat{z}$ and can never touch.

One can also consider the corresponding evolution of the Berry
curvatures and Chern transfers for finite slabs, where the bulk
of the slab undergoes the same cyclic evolution.  If the surface
Hamiltonian could be constantly readjusted so as to remain insulating,
the net result at the end of the cycle would be to change the
surface AHC by $\pm e^2/h$ at the bottom
and top surfaces of the slab respectively.  In the more common case
that the surface returns to its initial state at the end of the
cycle, the AHC must return to itself too, so the slab is topologically
required to have a metallic surface phase over some interval of
$\alpha$, such that the extra Chern number can be removed.
The existence of such surface states can be an experimental signature
characterizing any adiabatic loop with non-zero second Chern number.
  
In summary, we have demonstrated that an analysis of the WCC sheets
as defined in the HWF representation, which had previously been shown to
be useful for identifying and visualizing the topological
properties of non-trivial insulating phases, also provides an
insightful characterization of a non-trivial parametric loop
characterized by a second Chern number.  By defining Berry connections
and curvatures associated with the WCC sheets, we have derived
a new formula for the CSA axion coupling $\theta$, emphasizing that
it is naturally decomposed into a topological Berry curvature dipole
term and a non-topological correction term.  We showed that the
WCCs exhibit a non-trivial zigzag behavior as a function of the
loop parameter.  However, unlike for a Chern or $Z_2$-odd insulator,
this zigzag evolution of the WCCs describes not the pumping of
charge, but the pumping of layer Berry curvature defined in terms
of the HWFs.  In our formulation the $2\pi$ ambiguity of $\theta$
is readily evident when some sheets have non-zero Chern numbers,
in which case a different assignment of sheets to the home unit
cell can shift $\theta$ by $2\pi$,
and the link to the surface anomalous Hall conductivity becomes more
direct.  We also speculate that Eqs.~(\ref{eq:theta_zo}-\ref{eq:theta_dxy})
may provide a more efficient practical means of computing $\theta$
than those used previously, since there is no need to establish a smooth
gauge in the 3D Brillouin zone.
In any case, we believe that our extended development of the HWF
representation should prove broadly useful in characterizing
the adiabatic evolution of topological materials and their
magnetoelectric properties.

This work was supported by NSF Grant DMR-14-08838, and was
inspired in part by preliminary calculations of Sinisa Coh.
We thank Ivo Souza for useful discussions.


\bibliography{chern_pump}

\end{document}